\documentclass[conference,compsoc]{IEEEtran}
\usepackage[frozencache,cachedir=.]{minted}
\usepackage{fullpage}
\usepackage{times}
\usepackage[normalem]{ulem}
\usepackage{ dsfont }
\usepackage{ amssymb }
\usepackage{fancyhdr,graphicx,amsmath,amssymb, mathtools, scrextend, titlesec, enumitem}
\usepackage[vlined,linesnumbered,ruled]{algorithm2e}
\usepackage{changepage}
\usepackage{graphicx}
\usepackage{minted}
\graphicspath{ {./images/} }
\usepackage{hyperref}
\hypersetup{
  colorlinks=true,
  citecolor=magenta,
  linkcolor=red,
  urlcolor=blue,
}

\usepackage[
backend=biber,
style=ieee,
]{biblatex}
\usepackage{dblfloatfix}

\begin{document}

\title{The Fuse XORier Lookup Table \\ \large Exploration, Implementation, and Revision of Probabilistic Sets and Maps
}

\author{\IEEEauthorblockN{Eric Breyer}
\IEEEauthorblockA{School of Computer Science\\ Rice University '25\\ Email: eab17@rice.edu \\ \href{https://github.com/ericbreyer}{https://github.com/ericbreyer}}
\and
\IEEEauthorblockN{Alan Liu}
\IEEEauthorblockA{School of Computer Science\\ Rice University '25\\
Email: ayl8@rice.edu \\ \href{https://github.com/alanyliu}{https://github.com/alanyliu}}}
\IEEEspecialpapernotice{(Comp 480 Term Project)}

\maketitle

\begin{abstract}
This paper presents an exploration, implementations, and revisions of probabilistic sets and maps, specifically focusing on Bloomier filters and related data structures. The paper introduces the Fuse XORier Lookup Table (FXLT), an enhanced version of the Bloomier Filter incorporating spatial coupling, linear construction, and optimizations. The authors provide implementations in C and Python, comparing the FXLT's performance with other data structures like bloom filters, XOR filters, binary fuse filters, hash tables, and red-black trees. The FXLT demonstrates improvements in both space and time efficiency over traditional Bloomier Filters and appears competitive with hash tables for large datasets.
\end{abstract}

\section{Introduction}
Chazelle et al. (2004) designed a generalization of Bloom filters that could associate a value with each element that had been inserted, implementing an associative array \cite{Chazelle_All_2004}. Like Bloom filters, these structures achieve a small space overhead by accepting a small probability of false positives. In the case of "Bloomier filters", a false positive is defined as returning a result when the key is not in the map. The map will never return the wrong value for a key that is in the map. The ideas that they used have since been applied to sets as well with the XOR filter \cite{Graf_Lemire_2020}.

There has been some research into extensions of the XOR filter related to spatial coupling when building the table \cite{Graf_Lemire_2022}, which involves splitting the filter into segments for more fine-grained hashing (speedup or cache locality benefits). We bring these ideas back to the original Bloomier Filter implementation, augmenting the construction algorithms for faster construction, both in asymptotic complexity and constant reductions, and lower memory requirements. 

In summary, our goal is to profile existing probabilistic set data structures in order to create an associative array with sublinear space requirements, compact representation on disk/over network transfers, and tractable time complexity in construction and search. 
\section{Motivation}
The motivation for an associative array is a natural extension of the use cases of current filters. For example, using a Bloom or XOR filter we can encode a set of malicious URLs, ask if a URL is malicious, and get an answer of “definitely not malicious,” or “probably (with high likelihood) malicious.” However, what if a user is okay with some level of risk? What if they want to only be warned about some risk levels but not others, or even change that risk level? An encoding for an associative array would allow us to ask if a URL is malicious, and get an answer of “definitely not malicious”, “probably not secure but not malicious”, “probably dangerous to a naive user”, “probably actively harmful”, etc. When a yes/no answer won’t do, sets leave us out of luck.
\section{Previous Work}
This paper assumes familiarity with the Bloomier Filter \cite{Chazelle_All_2004}. Passing familiarity with the XOR Filter and Binary Fuse Filter \cite{Graf_Lemire_2020}\cite{Graf_Lemire_2022} would also be helpful.
\begin{table*}[!t]
\renewcommand{\arraystretch}{1.3}    
\caption{Literature Survey}
\label{tab:Literature Survey}
\centering
\begin{tabular}{|c|c|c|c|c|}
    \hline
    Structure (k = num hash funcs) & Purpose & Dynamic & Construction Time & Space Requirement \\
    \hline
    Bloom Filter (k = optimal)& set & supports insertions (to a point)&O(n) & $1.44n * log_2(\epsilon^{-1})$\\
    Bloomier Filter (k = 3)& map & static & O(nlogn) (avg) / O($n^2$) (worst)&$1.23n * log_2(\epsilon^{-1})$\\
    XOR Filter (k = 3)& set &static & O(n)&$1.23n * log_2(\epsilon^{-1})$\\
    Binary Fuse Filter (k = 3)& set &static & O(n)&$1.125n * log_2(\epsilon^{-1})$\\
    Binary Fuse Filter (k = 4)& set &static & O(n)&$1.075n * log_2(\epsilon^{-1})$\\
    \hline
\end{tabular}
\end{table*}

\subsection{Bloomier Filter \cite{Chazelle_All_2004}}
This structure is arguably misnamed as it is very loosely related to Bloom Filters. They both use multiple hash functions indexing into an array, but whereas Bloom Filters use an “AND” scheme to determine membership, Bloomier Filters use an “XOR” scheme. This leads to many differences. 
\begin{enumerate}
    \item Bloom Filters support insertion, whereas Bloomier Filters are static after construction
    \item Bloomier Filters have space requirements closer to the theoretic minimum ($1.23n * log_2(\epsilon^{-1})$ vs $1.44n * log_2(\epsilon^{-1})$ for bloom filters)
    \item Insertion into a bloom filter is O(1), so construction given an input set of n elements is O(n) while construction of a Bloomier filter is O(nlogn)
    \item Bloomier filters can associate items with more fine-grained values while Bloom filters can only answer binary membership queries
\end{enumerate}
\subsection{XOR Filter \cite{Graf_Lemire_2020}}
In 2020, Thomas Graf and Daniel Lemire theorized a data structure heavily based on the Bloomier Filter known as the XOR filter. The XOR filter is essentially an implementation of a Bloomier Filter for sets, which is a trivial change - instead of storing indexes into an array we store unique fingerprints. However, important for us, Graf and Lemire provide a more efficient construction algorithm than was shown in the previous paper. With an efficient implementation of the algorithm, construction can be accomplished in O(n) time (vs the original O(nlogn) construction). 
\subsection{Binary Fuse Filter \cite{Graf_Lemire_2022}}
In 2022, Graf and Lemire outdo themselves with the Binary Fuse Filter. Again, this is an unnecessarily confusing name, the structure is really just a XOR filter with better construction, allowing for faster construction and, more importantly, lower space requirements. The construction uses a technique known as spatial coupling. Spatial coupling is an augmentation to the “peeling” process of the Bloomier and XOR filters, and restricts the hash functions to localized consecutive “windows” instead of letting them access the whole array. This typically results in less hash collisions at the ends of the array, allowing those items to be peeled first, which in turn frees up the next outermost layers and so forth. This is how the structure gets its name: the peeling process looks like a fuse burning from the outside in. Less collision probability = Less space needed.
\subsection{Survey}
Since the original paper on bloom filters, there has been progress on their use as sets, but these learnings have not been applied back to the Bloomier filter. Additionally, there is still very little interest in implementing and profiling these structures. We aim to rectify both of these issues. See Table \ref{tab:Literature Survey} for an overview of the current literature.

\section{Argument}
We argue that the original construction of a probabilistic associative array is outdated. By applying the spatial coupling technique of the Binary Fuse Filter, the linear construction of the XOR filter, optimizing the algorithm for real-world implementation, and giving it a better name, we make a sublinear space and time-efficient associative array known as the \emph{Fuse XORier Lookup Table}, hereby catapulting the Bloomier Filter into the 21.2nd century. 

There are currently no tractable implementations and a lack of real-world tests for probabilistic associative arrays. We address this by providing implementations in C (w/ wappers for cpp) and python.
\section{Fuse XORier Lookup Table}
The Fuse XORier Lookup Table (FXLT) is an augmentation of the Bloomier Filter that incorporates Spatial Coupling, a Linear Construction Algorithm, and real-world optimizations. Note: this structure still has the same false positive probabilty as the Bloomier Filter: $\epsilon = 2^{-b} * k$ where $b = $ bits per element. As such $b \sim log_2(\epsilon^{-1})$, and the size of the FXLT will be $cnb \sim log_2(\epsilon^{-1})$ for some constant $c$ and $n$.

\begin{figure*}[bpb]
    \centering
    \smaller
    \caption{Bloomier Filter Hashing}
    \label{alg:naive-hash}
    \begin{minted}[mathescape]{c}
void bloomierFilter_hashAll(struct BloomierFilter *self,
                                SizedPointer elem, size_t *neighborhood) {
// the k location hash values
for (size_t i = 0; i < self->k; ++i) {
    neighborhood[i] =
        murmur3_32(elem.ptr, elem.size, self->hash_seeds[i]) % self->m;
}
// the random M value
neighborhood[self->k] =
    murmur3_32(elem.ptr, elem.size, self->hash_seeds[self->k]) &
    (((uint64_t)1 << self->q) - 1);
}
\end{minted}
\end{figure*}

\subsection{Spatial Coupling}
The spatial coupling technique allows us to reach lower bounds on space requirements by augmenting the construction method. The original Bloomier Filter algorithm hashes elements into the range of the entire filter, but if we instead make the locations an element hashes to close together, we make it far less likely for there to be collisions at the edges of the filter. Intuitively, this is because less "windows" overlap with the very first or very last elements, and thus the expected number of items hashing to these locations is lower. Thus the chance of items on the edges being singletons is much higher, meaning they will get "peeled" first, which will in turn expose new singleton locations. Once again, the way items get peeled looks like a double ended fuse, hence the name.

\subsubsection{Construction}
The algorithm used to hash elements in the Bloomier Filter is given in figure \ref{alg:naive-hash}. This algorithm simply hashes the element $k$ times within the range of $m$ (the number of slots in the table) along with a special hash value used for randomization.

In a Fuse XORier Lookup Table, we use spacial coupling technique in an analogous way to the Binary Fuse Filter. We choose an optimal segment size $w$ based off the number of hash functions as given in the Binary Fuse Filter paper: $4.8n^{.58}$ for $k=3$ and $.7n^{.65}$ for $k=4$. Instead of hashing into the entire range of $m$, we will select three distinct locations within three consecutive segments. To accomplish this, we will pick the segments first and then hash the element into the range of these segments. This algorithm requires $k+2$ hash functions. First, we use a hash function to get a random segment index. Segments have size $w$, so $firstSegmentIdx * w$ is the start of our first segment. To get the start of the segment corresponding to hash function $i$, we just move one segment size ($w$) over: $firstSegmentIdx * w + w * i$. Segments have size $w$ so the range of the segment corresponding to hash function $i$ with $firstSegmentStart = firstSegmentIdx * w$ is $[firstSegmentStart + i * w, firstSegmentStart + i * w + w)$. With this, we just hash each element into the hash function's computed segment, and of course get out random $M$ value. This algorithm is given in figure \ref{alg:fxlt-hash}.

\begin{figure}[H]
    \caption{Spacial Coupling Space Advantage}
    \label{fig:CoupledSpaceAdv}
    \centering
    \includegraphics[width=0.39\textwidth]{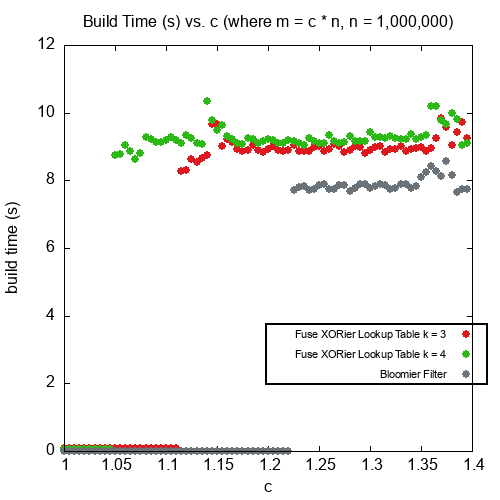}
\end{figure}
\subsubsection{Performance}
\begin{figure*}[tp!]
    \caption{Buildable Boundary for $k=4$}
    \label{fig:BuildBoundary}
    \centering
    \includegraphics[width=.87\textwidth]{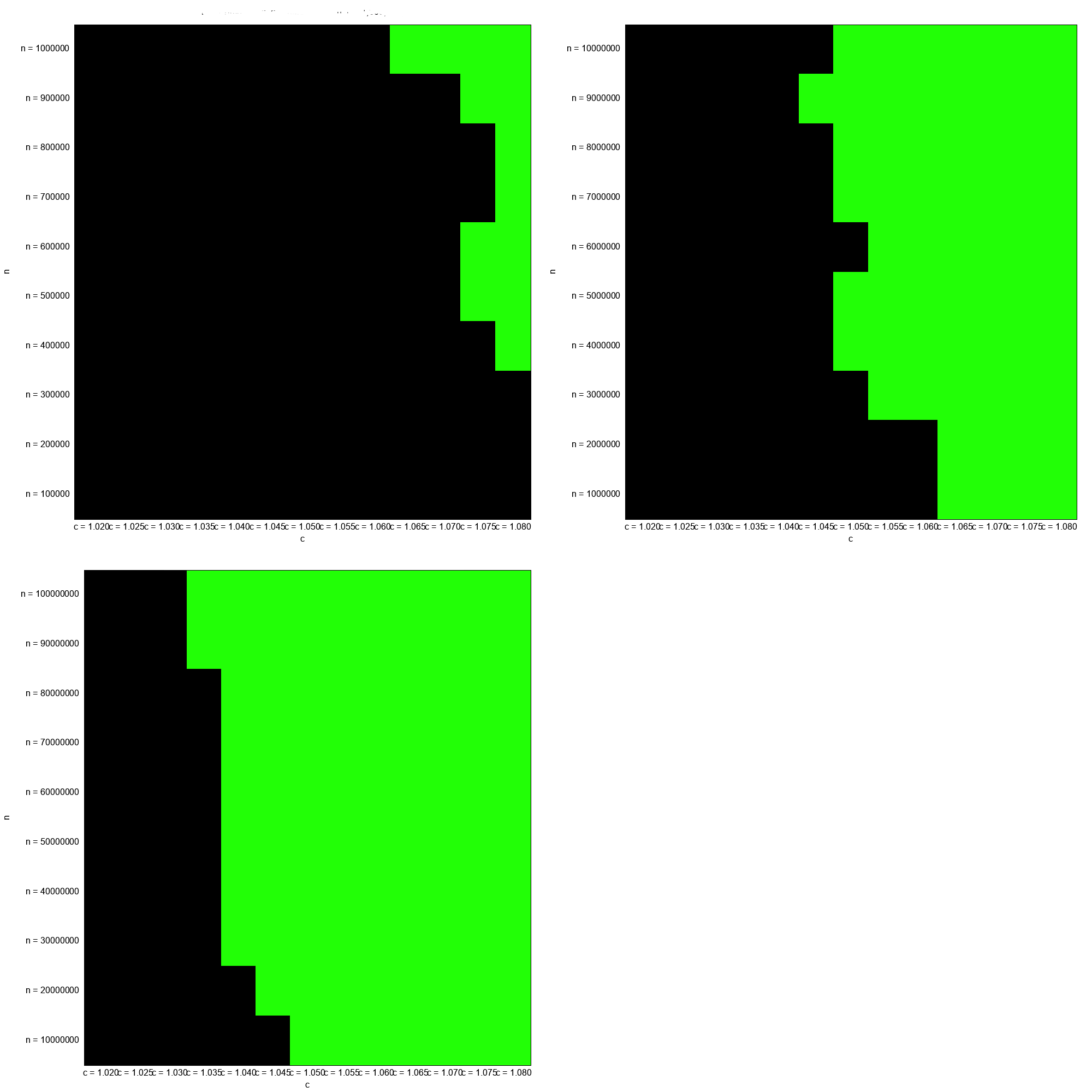}
\end{figure*}

Figure \ref{fig:CoupledSpaceAdv} plots build time with respect to space for a Bloomier Filter and Fuse XORier Lookup Tables with k=3/k=4 and optimal window size. We see that the more hash functions are used, the more time construction takes, but this is to be expected. Points at 0 denote that the filter failed to build at that value of c. Thus we see that the Fuse XORier Lookup Table achieves lower memory requirements than a Bloomier Filter. The Bloomier Filter has a hard limit at $m = 1.23n$, but FXLTs can reach down lower. For memory usage, $k=4$ hash functions provides a better bound, (experimentally $c = 1.3$) than $k = 3$ ($1.08$) at the expense of some build time. Figure \ref{fig:BuildBoundary} shows the experimental memory bound for a $k=4$ FXLT. Black means the filter failed to build at those values of $c$ and $n$. We get closer and closer to the bound the bigger $n$ gets, but the limit is around $m = 1.03n$ ($1.03$ slots per element) when $n$ is large. This is a huge improvment over the Bloomier Filter at $c = 1.23$ ($1.23$ slots per element) and even the Bloom Filter with $c = 1.44$ ($1.44$ slots per element). The XOR strategy with fuse construction puts us very close to the theoretical lower bound of $c = 1$.

\begin{figure*}
    \centering
    \smaller
    \caption{FXLT (spacial coupling) Hashing}
    \label{alg:fxlt-hash}
    \begin{minted}[mathescape]{c}
void fuseXORierLT_hashAll(struct fuseXORierLookupTable *self,
                                SizedPointer elem, size_t *neighborhood) {
// get the first window
size_t firstSegmentIdx = murmur3_32(elem.ptr, elem.size, self->hash_seeds[self->k + 1]) 
                          % (self->numWindows - self->k)

size_t firstSegmentStart = firstSegmentIdx  * self->w;

for (size_t i = 0; i < self->k; ++i) {
    size_t segmentStart = firstSegmentStart + i * self->w;
    size_t temp = murmur3_32(elem.ptr, elem.size, self->hash_seeds[i]);
    neighborhood[i] = temp % self->w + firstSegmentStart;
}
neighborhood[self->k] =
    murmur3_32(elem.ptr, elem.size, self->hash_seeds[self->k]) &
    (((uint64_t)1 << self->q) - 1);
}
\end{minted}
\end{figure*}

\begin{figure*}[b]
    \centering
    \smaller
    \caption{XOR Set}
    \label{alg:XOR-Set}
    \begin{minted}[mathescape]{c}
struct xorSet {
    size_t elems;
    size_t cardinality;
};

void addElem(struct xorSet *set, size_t elem) {
    set->elems ^= elem;
    ++set->cardinality;
}
void removeElem(struct xorSet *set, size_t elem) {
    set->elems ^= elem;
    --set->cardinality;
}
bool singleton(struct xorSet *set) { 
    return set->cardinality == 1; 
}
size_t getSingleton(struct xorSet *set) { 
    return set->elems; 
}
\end{minted}
\end{figure*}

\subsection{Linear Construction}
Bloomier Filters use a construction algorithm that requires the set of singleton slots to be recalculated every iteration until there are either no singletons or all items are places \cite{Chazelle_All_2004}. This requires iterating through all unplaced elements every iteration, resulting in an $O(nlogn)$ running time on average (assuming the number of unplaced elements decreases by a constant factor each iteration). 

However, we can do better \cite{Graf_Lemire_2020}. First, build an array of sets where index $i$ stores the set of keys that hash to location $i$. Then, initialize a queue with every location in the array that is a singleton set. We then loop until this queue is empty. When we process a singleton, we remove it from all sets of locations it hashes to in the array. We then check if any of those sets become singletons as a result and add them to the queue. Since we use a queue, the required ordering property of the algorithm holds - the first elements popped off the queue can be placed last because no later element will rely on that slot (the slot is singleton). The loop runs until the queue is empty, and given space requirements equivalent to that of a binary fuse filter with spatial coupling, all $n$ elements will be pushed onto the queue. Each loop iteration is $O(1)$, so this algorithm is $O(1)$. 

In practice the large number of sets we need could slow us down, especially if we use something heavyweight like hash tables. However, since we only ever need to know what the id of an element that hashed to a slot was if the slot is a singleton, these sets can be efficiently implemented as a counter and a bitmask initialized to 0 \cite{Graf_Lemire_2020}. To insert into a "XOR set", we increment the counter and set the bitmask to the XOR of the item with the bitmask. Similarly to remove an element we decrement the counter and set the bitmask to the XOR of the item with the bitmask. To check if a set is a singleton, check if the counter is $1$. If it is, the value of the bitmask is the value of the element. See figure \ref{alg:XOR-Set} for a code sample. Using this set implementation drastically reduces constant factors on auxiliary memory and run time.
\subsubsection{Performance}
\begin{figure*}[t]
    \centering
    \caption{Build Time Old vs New Construction}
    \label{fig:oldBuildTime}
    \includegraphics[scale=.42]{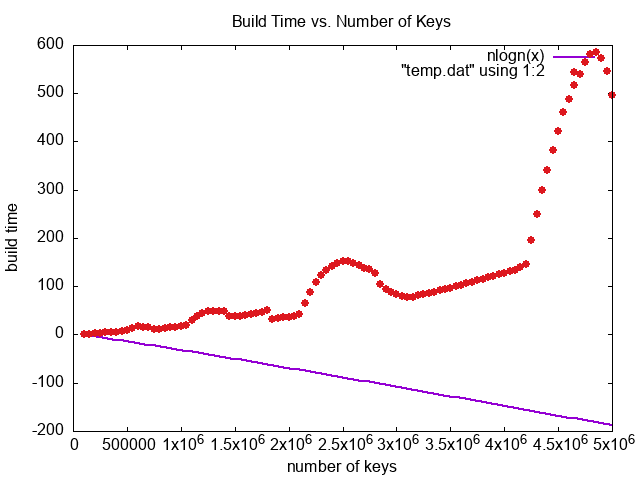}
    \includegraphics[scale=.42]{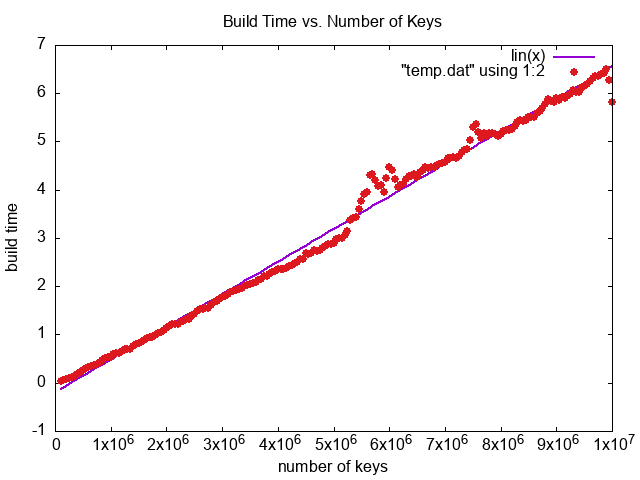}
\end{figure*}

Figure \ref{fig:oldBuildTime} shows the build time vs number of keys for the old construction on the left and the new construction on the right.
We can see the linear vs log linear difference, and also a consistency difference. Most dramatically however, are the running times themselves. The old construction takes minutes to construct a filter of size $5 \cdot 10^6$ while the new construction can construct a $10^7$ size lookup table in mere seconds. $n=10^8$ filters are also tractable, and the limiting factor for building an $n=10^9$ filter is not being able to hold all the keys in RAM during construction (on a 16GB M1 MacBook Air).

\begin{figure*}[t]
    \centering
    \caption{Cached vs Non-Cached Construction}
    \label{fig:cacheornah}
    \includegraphics[width=.84\textwidth]{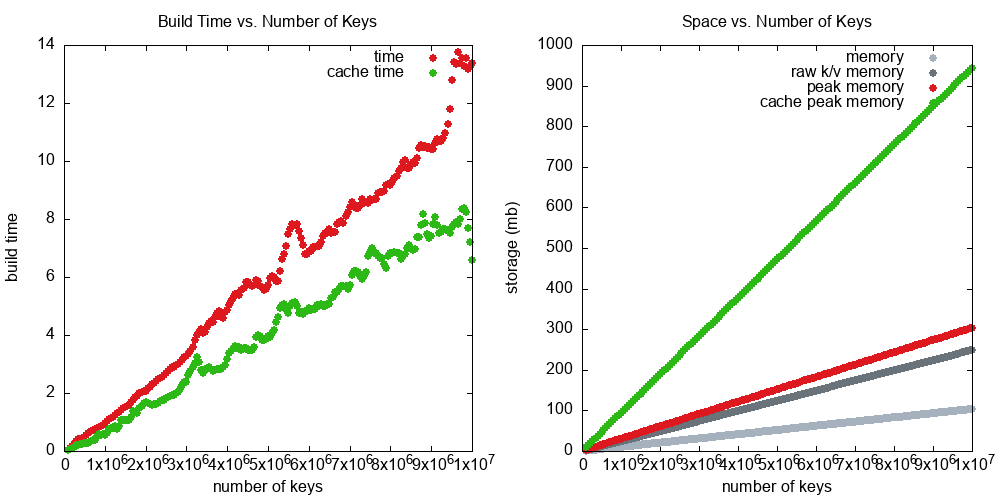}
\end{figure*}

\subsection{Caching Hashes}
Although the linear algorithm reduces the number of times a key must be hashed, there is still redundant hashing every time we need to find a key's location. In construction this happens 3 times. If we cache the hash values of each key, construction should be faster at the expense of some auxiliary memory (denoted as peak memory in the graph). We also show the final memory of the constructed FXLT and the memory of the keys and values themselves (a hash table has to store keys and values so this is like a hash table lower bound). Figure \ref{fig:cacheornah} shows the trade off in both run time and memory. Caching is a flag in construction and as such can be turned on or off depending on the user's needs. Caches are discarded after construction, and as such do not effect the final memory footprint.
\begin{figure*}[b]
    \centering
    \caption{Hash Table Performance Comparison }
    \label{fig:hashtablecompar}
    \includegraphics[width=.85\textwidth]{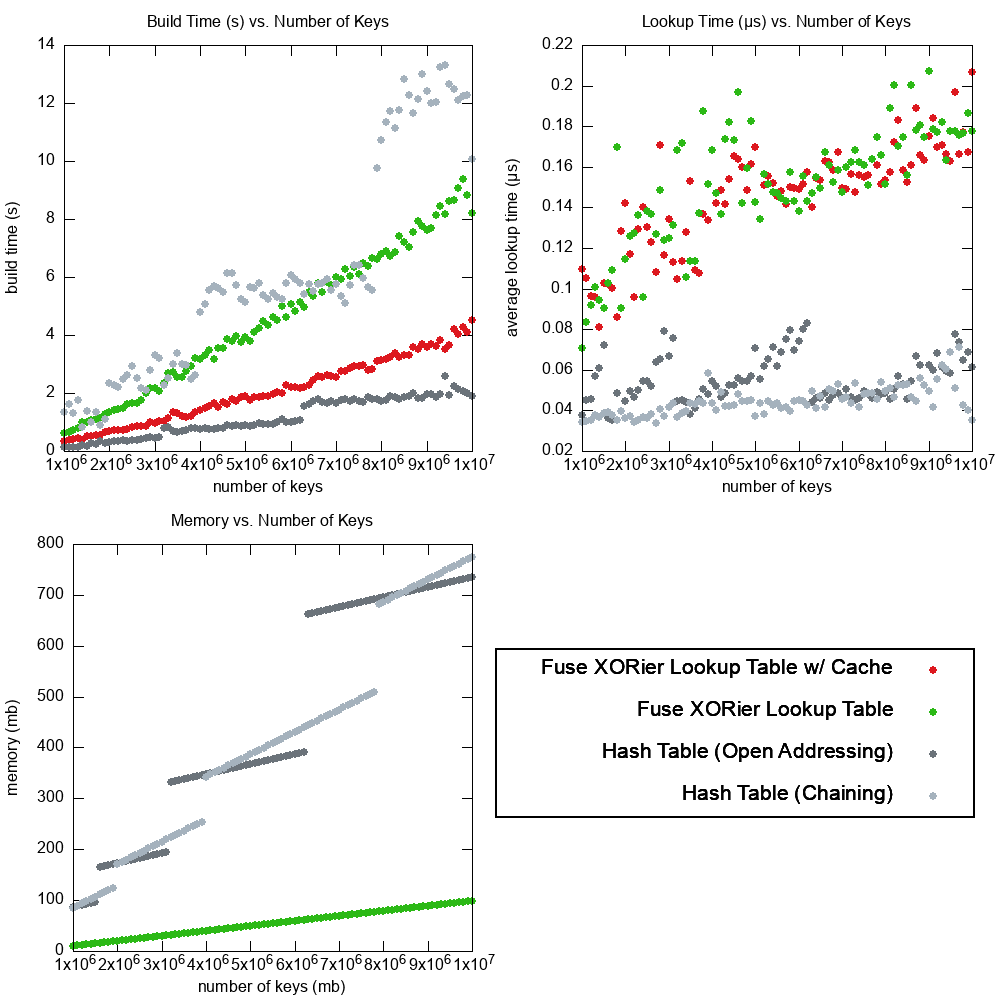}
\end{figure*}
\begin{table*}[t]
\renewcommand{\arraystretch}{1.3}    
\caption{The FXLTs Place in the world}
\label{tab:Literature Survey w/ FXLT}
\centering
\begin{tabular}{|c|c|c|c|c|}
    \hline
    Structure (k = num hash funcs) & Purpose & Dynamic & Construction Time & Space Requirement \\
    \hline
    Bloom Filter (k = optimal)& set & limited insertions&O(n) & $1.44n * log_2(\epsilon^{-1})$\\
    Bloomier Filter (k = 3)& map & static & O(nlogn) (avg) / O($n^2$) (worst)&$1.23n * log_2(\epsilon^{-1})$\\
    XOR Filter (k = 3)& set &static & O(n)&$1.23n * log_2(\epsilon^{-1})$\\
    Binary Fuse Filter (k = 3)& set &static & O(n)&$1.125n* log_2(\epsilon^{-1})$\\
    Binary Fuse Filter (k = 4)& set &static & O(n)&$1.075n* log_2(\epsilon^{-1})$\\
    \hline
    XORier Lookup Table (No Spatial Coupling) & map & static & O(n) &$1.23n* log_2(\epsilon^{-1})$\\
    Fuse XORier Lookup Table (k = 3) & map & static & O(n) & $1.125n* log_2(\epsilon^{-1})$\\
    Fuse XORier Lookup Table (k = 4) & map & static & O(n)  & $1.075n* log_2(\epsilon^{-1})$\\
    \hline
    Hash Table & map/set & fully & O(n) & O(n) (stores the full key)\\
    Balanced BST (and Skip List) & map/set &fully &  O(nlogn) & O(n*keysize) (stores the full key) \\
    \hline
\end{tabular}
\end{table*}

\section{Relative Performance}
We have shown that the Fuse XORier Lookup table outperforms the Bloomier Filter, but we will now show its efficacy compared to more traditional data structures used to implement associative arrays: namely hash tables (using chaining and open addressing) and Red-Black Trees (compare to other Balanced BSTs and Skip Lists). FXLTs are more specialized than these other implementations. Hash Tables and Balanced BSTs are dynamic, supporting insertions and deletions, whereas FXLTs are static upon construction. Additionally, FXLTs have a small chance to report that an element is in the map with a bogus value, while Hash Tables and Balanced BSTs do not. In exchange for this smidgen of error and its static nature, FXLTs offer an encoding in \emph{sublinear} space. The size of the structure is less than the size of the keys and values that were inserted.

We compare these structures on three metrics: 
\begin{enumerate}
    \item Build Time - this measures the time taken to build a structure containing $n$ keys (or time for $n$ insertions in the case of the dynamic structures)
    \item Lookup Time - this measures the time taken to lookup a single item in the structure.
    \item Memory - the memory footprint of the built structure.
\end{enumerate}

Figure \ref{fig:hashtablecompar} Shows the FXLT with and without cache compared to Hash Tables using Open Addressing and Chaining for collision resolution. 

Starting with build time, we see that all structures are build in asymptotically $O(n)$ time with respect to the number of keys. This is expected and good.
We can see that the FXLT takes longer to build that the Open Addressing Hash Table which makes sense. The FXLT has some extra bookkeeping in order to maintain linear construction time mentioned in 5.2 for which the hash table doesn't have to deal with. Interestingly, the chaining hash table is worse than even the FXLT without caching. This may just be because my implementation isn't optimised, or that the chaining hash table has to deal with more overhead related to cache misses and dynamic memory management. Overall, the FXLT is not orders of magnitude slower than a open addressing hash table, so it should be usable.

In lookup time we see the hash tables do better than the FXLTs. This is again to be expected, as FXLTs have 4 or 5 more hash functions than the hash tables, which is the main bottleneck during look ups. Interestingly, the FXLT lookups look to be asymptotically $O(logn)$. We would expect this to be $O(1)$ (or more accurately $O(k)$), so this is weird behavior. I suspect it has something to do with the modulo operations to keep the hash values in range, or some kind of cache misses with more memory. However, these values are all still on the order of nanoseconds, and are less than $2x$ a hash table, so no worries with lookup time.

Lastly, we turn to memory. This is where the FXLT is meant to shine, and we clearly see that it does. All structures are $O(n)$ in memory usage, but since the FXLT does not store the keys itself, it is able to use orders of magnitude less memory. 

\begin{figure*}[b]
    \centering
    \caption{Balanced BST Performance Comparison}
    \label{fig:rbtreecompar}
    \includegraphics[width=.8\textwidth]{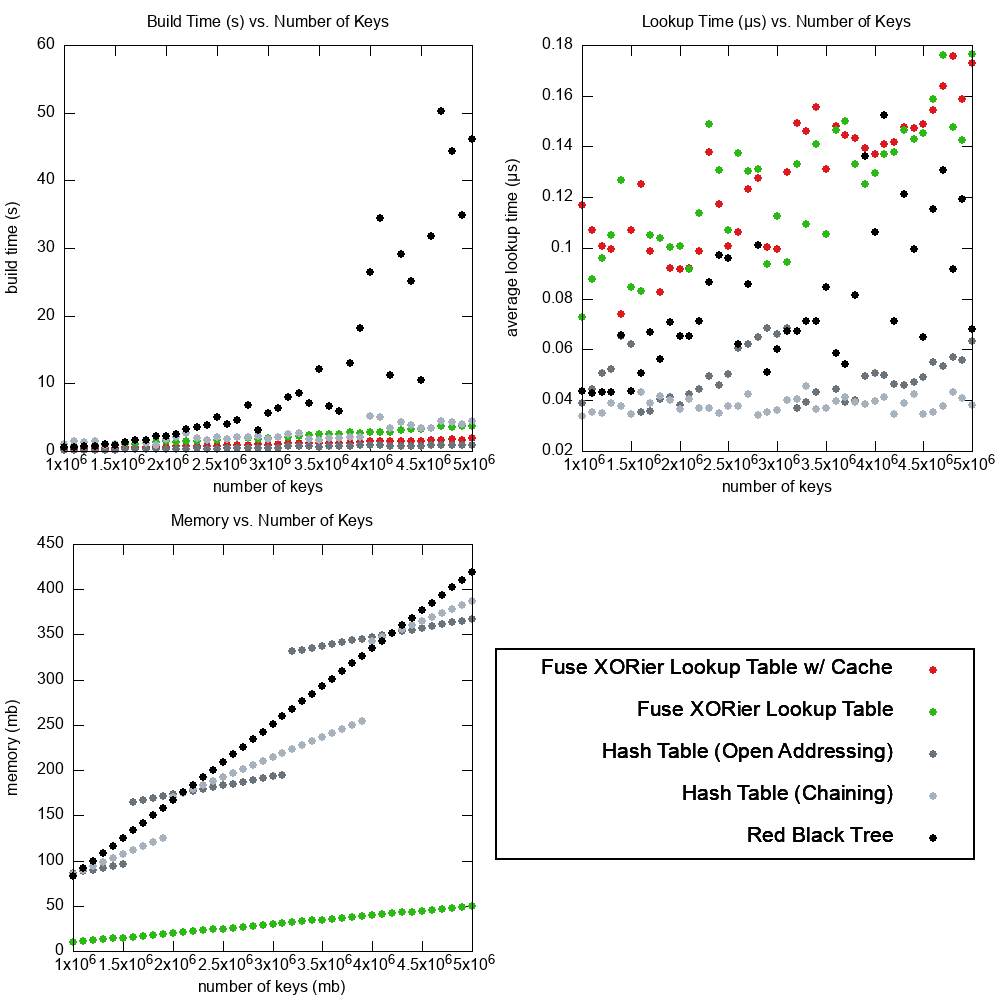}
\end{figure*}

Figure \ref{fig:rbtreecompar} adds a Red Black Tree into this comparison. A Balanced BST has very different asymptotic complexity so the comparison is not as useful. In fact, the Build Time for the Red Black Tree is divided by $100$ in order to fit anything on the plot.

\section{Conclusion}
Table \ref{tab:Literature Survey w/ FXLT} adds the FXLT to the literature survey. The FXLT improves on the Bloomier Filter by incorporating advancements from the XOR filter and Binary Fuse Filter \cite{Chazelle_All_2004}\cite{Graf_Lemire_2020}\cite{Graf_Lemire_2022}. It acheives great performance advantages in space and time over a standard Bloomier Filter, and is usable compared to a hash table even for very large numbers of entries. If one can sacrifice some false positive rate, the Fuse XORier lookup table can greatly improve performance with it's encoding in sublinear space. 

\section{Code}
\begin{itemize}
    \item C Implementation (Eric Breyer) - 
    \href{https://github.com/ericbreyer/C-Fuse-XORier-Filter}{Github Link}
    \item Python Implementation (Alan Liu) -\href{https://github.com/alanyliu/python-fuse-xorier-filter}{Github Link}
\end{itemize}

\newpage
\nocite{*}
\printbibliography

\end{document}